# Examining the Efficiency of Sun Lightening and Shadow Tools in AutoCAD Program

Emad Ali Al-Helaly
Kufa University
College of Engineering

Israa Jameel
Baghdad University
College of Science

## Abstract:

The AutoCAD is one of the most famous Computer Aided Drawing programs with high and accurate specifications in the engineering design. It is highly qualified and contains many of the tools it needs in many engineering departments.

A useful tool is the lighting tool because it gives a simulated rendering of reality to a great degree that benefits the architect as well as urban designers. This tool includes simulating sunlight by date during the year, timing of day, and position of the body on the Earth.

We tested the sunlight status tools in this program in the limit of our test region to see how accurate it was and it turned out to have a 45% error difference. The sun shadow in AutoCAD rendering is longer than the real by 145%. There is another error in the direction of shadow also.

It is essential to note these errors for any designer need to calculate the shadow length and direction from the AutoCAD program.

## 1. Introduction: light and shadow

The light is the electromagnetic rays that coming from sun in the visible range. They are propagate in straight line always. Therefore, they cannot turn about the opaque bodies except in small ratio due to refraction near the edge of the bodies.

The shadow is the darkness that caused by an object when it blocks the light from getting a corresponding surface [R.1].

In practice, the (shadow) expression is a term used for referring to the areas behind objects where the light has blocked by that object, the shadow area is always opposite to the light source, and in opposite direction. In addition, if the light source is spread out and scattered the shadow does not have sharp edges and it is not nearly as defined as in the case when the sky is cloudy. The shadow occurs in the image when there is an object that blocks the light coming from the light source partially or completely [R. 2].

## 2. Simulating of sun light and shadow graphically

In any model simulates the sun light, we need to put a light source simulates the sun. The sun shadow casted by any body under the sun depends on the sun orientation, if there is another body away from the first body the different in shadow direction between them depends on the radial angle that is formed from the two line between the body and the sun. This case need to design a virtual studio simulate the a percentage of distances between the earth and the sun, which is so difficult, not practical, and need to give a high luminance to the created light in the virtual studio (3000 to 3500 Kelvin [R. 3]), some designer considered the sun as infinitely bright [R.4].

In the modern graphic programs it is seemed that the simulating of sun light and shadow made by putting a light source in a relatively far distance and then calculate the angle of shadow from that point approximately.

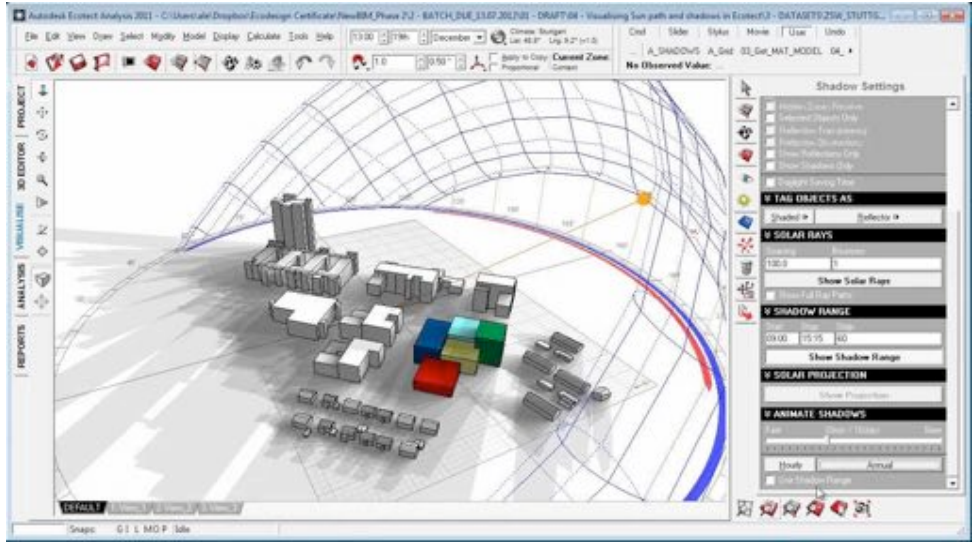

Figure (1) The Autodesk Ecotect windows, note that it depends a virtual path to the sun less than the percent of sun location to the earth.

This problem decreases the reliability of these programs in the purposes which need high accuracy in simulation.

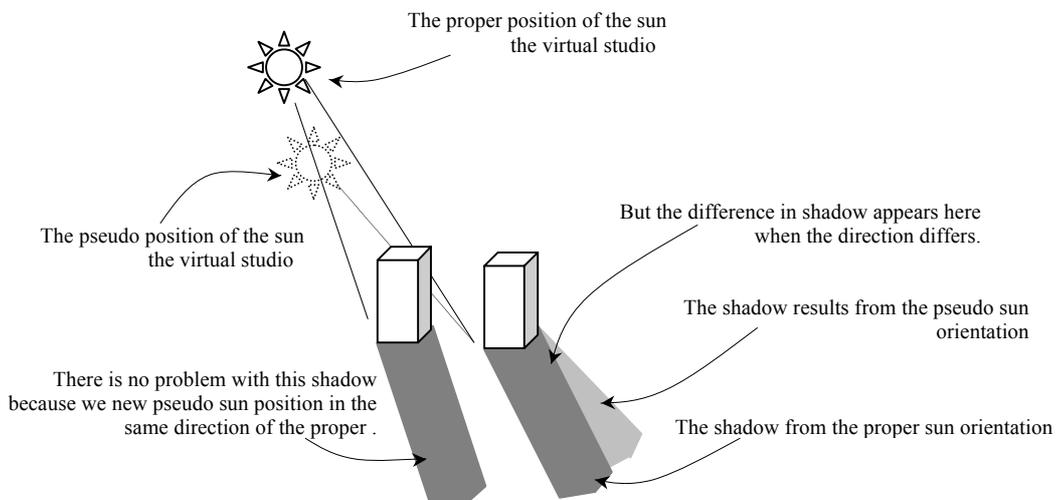

Figure (2) The difference in shadow according to the designed sun position.

The simulation of sun light is very difficult [R. 5] because of the many parameters affect the lighting of sun on the targets on the earth.

Kirk Tuck , 2009, suggested a diffused light source to simulate the sun brightness [R. 6] in the photography studio. It may be adequate to do in and software to simulate the brightness of sun and shadow to proper degree. But in the big targets like mountains and building it is not adequate procedure due to the direction of shadow.

Peter Tregenza and David Loe, 2009, suggested a complex calculation to simulate the sun light and path depending on the changing sun path and luminance [R. 7]. but they did not apply it in a computer model.

With the development of computer graphic hardware and software, the trying to simulate the lighting and rendering the natural lighting scene developed also.

W. Eric Wong and Tingshao Zhu, 2014, showed the simulating sunlight techniques in the modern software like 3D Max and some other programs [R. 8].

There are several software were developed in the two decays contain tools to simulate the sunlight and shadow for engineering, scientific, artificial purposes like Autodesk AutoCAD, Ecotect Analysis, Google Sketchup, Daylight Visualizer, 3D Studio Max, etc.

## 3. The geometry of shadow:

The shadow is usually a two-dimensional form that results from three-dimensional dark shapes and depends on its geometrical properties, the shape of the ground on which it falls, and at the angle of fall of the sun.

### The effect of the object shape on the shadow

The shadow section varies in its opacity and its edges depending on the various landmarks, for example, the tree shadow is different from the shadow of industrial structure even if it looks similar in its appearance.

Dark objects block all the wavelengths of the sun rays from the shadow area, so it will appears black, gray or close to black.

**Angle of the sun and its synchronization with space photography**

The sun is the main source of all types of electromagnetic radiation directly or indirectly, the sun appears moving in the sky along the day, and the shadow is changing according to the sun direction.

The angle of the sun is not constant for all the days of the year because of the difference in the angle of the fall of radiation on the ground. In addition, it is seemed moving in complex orbit on the sky from the observer on the Earth.

The incident angle of the sun light on the Earth's surface varies by distance from the equator because of the spherical shape of the planet. If we assume that the sun will be on March 21 perpendicular to the equator so that there are no shadows there, it will not be vertical in areas northern or southern equator. Here we will discuss in detail what is related to the angle of the sun and the earth, which affect the length, shadow and direction.

**The Sun Characteristics**

The sun's diameter is about 1.4 million km and the average distance between it and the Earth's surface is about 149.5 million kilometers. It produces huge nuclear reactions that produce various wavelengths of electromagnetic radiation ranging from long-range radios and short-wavelength cosmic rays. Electromagnetism consists of approximately 46% visible radiation and 46% thermal radiation [R. 9]

The Earth takes a biosphere orbit around the sun, so the amount of intensity falling on the ground varies throughout the year. The earliest distance to the Earth (147 million km) is at the beginning of January, and further beyond it (152 million km) at the beginning of July [R. 10].

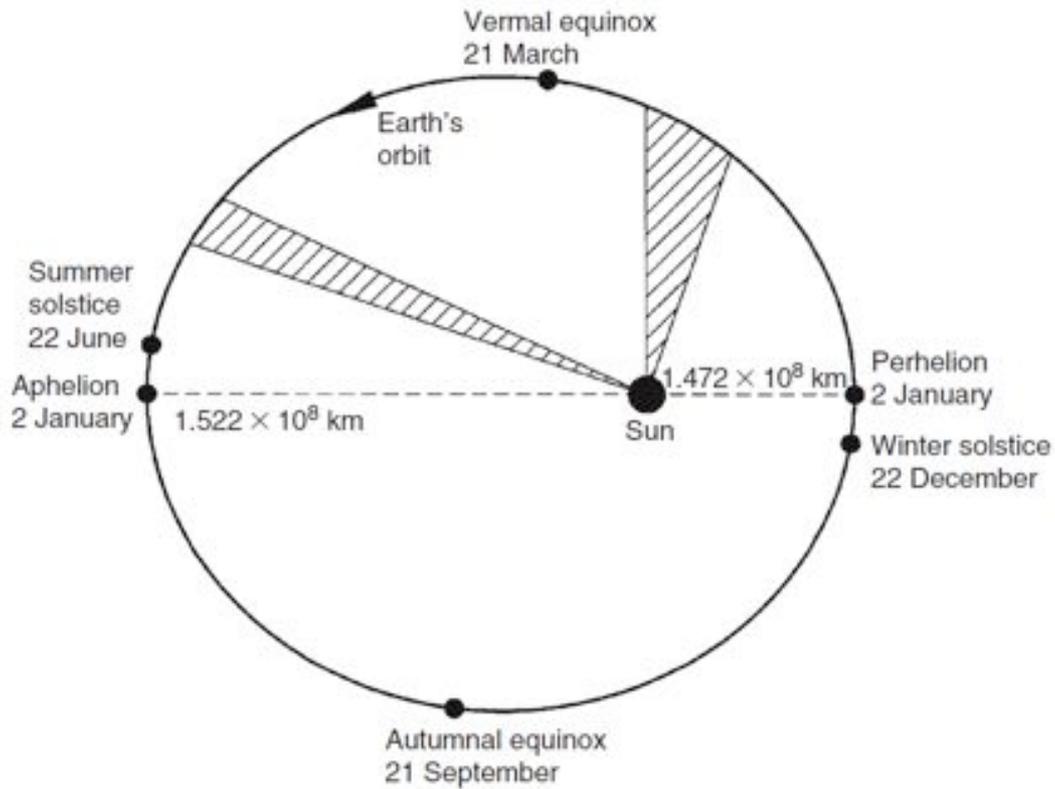

Figure (3): Earth's orbit around the sun [R. 11].

For more accurate calculations to the distance [R11, p. 3], let us take the mean distance between the sun and earth $(r_o)$= 1.496 ×$10^8$ km, or accurately 149597890±500 km [R. 11, p. 3], and (r) is the sun-earth distance for any day in the year. The day angle (Γ in radians) is:

Γ = 2π $(d_n$-1)/365 ……………………………………..(1)

Where: $d_n$ in the day number, which is 1 to the first of January, and 365 on the last day of December.

The Earth's axis of rotation across the year tends to be about ±23.5 degrees [R. 11], which affects the angle of the sun's fall and its difference on the surface of the earth, and thus affects the direction of the shadow and its extension.

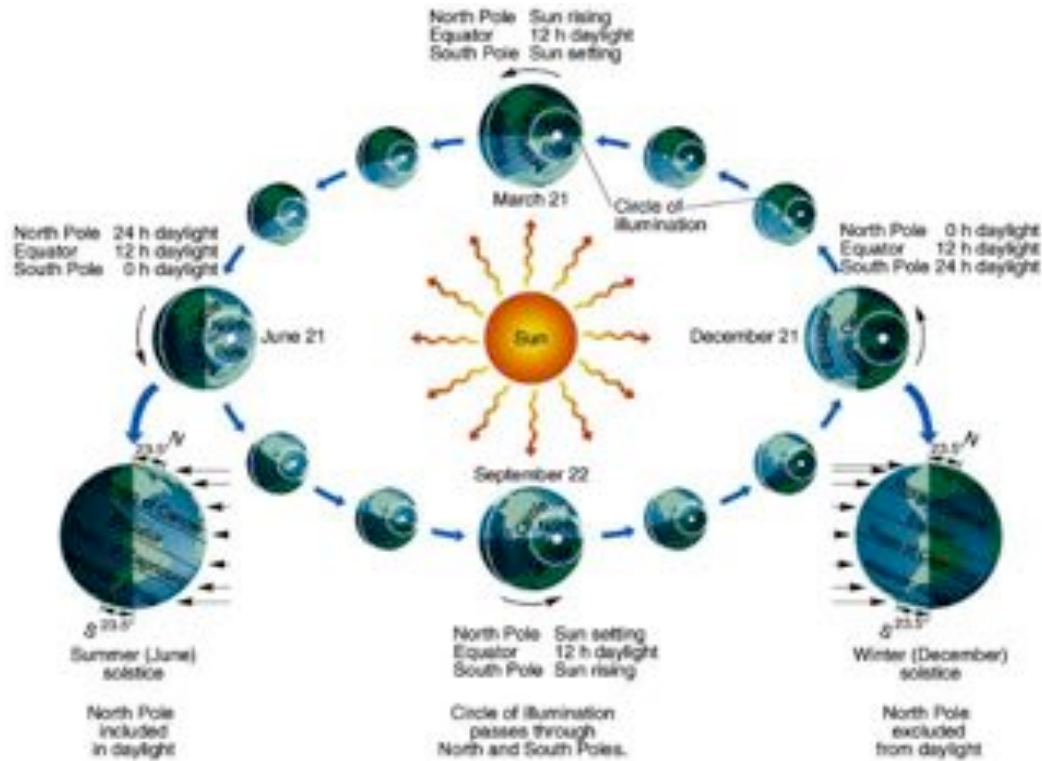

Figure (4): The difference in the axis of the Earth's
rotation around itself during the year [R. 10]

The angle, position and elevation of the sun can be calculated by means of theodolites. These calculations can be organized by astronomical equations [R. 11] in the following manner

**Sun location (angles of deviation, elevation, azimuth of sun)**

**The angle of solar deviation (δ):** it is the angle between the equatorial plane (the plane of earth revolution about the sun) and the line between the center of the Earth and the center of the sun. It is also called declination angle. The value of this angle ranges between +23.45 and -23.45, and the value of this angle varies from day to day. At 20/21 March and 22/23 September it is equals zero. It is calculated from the relationship:

$$\delta= (0.006918-0.399912\times\cos(\Gamma)+0.070257\times\sin(\Gamma)- 0.006758\times\cos(2\Gamma)$$
$$+0.000907\times\sin(2\Gamma)-0.002697\times\cos(3\Gamma)+0.00148\times\sin (3\Gamma) \times(180/\pi)$$

.....................................(2)

This equation is derived from a Fourier formula, was developed by J. W. Spencer [R. 12], and it is the most accurate equation. There are other simple equation like [R. 13]:

$$\delta = \pi \times 180 \, 23.45 \times \sin(2\pi \times 365 \, 284 + d_n) \ \dots\dots\dots\dots \ (3)$$

$d_n$: is the number of the day of the year as submitted.

Or this equation:

$$\delta = \sin^{-1} \times \sin(23.45^0) \times \sin(2\pi \times d_n - 81 \, 365) \dots\dots\dots(4)$$

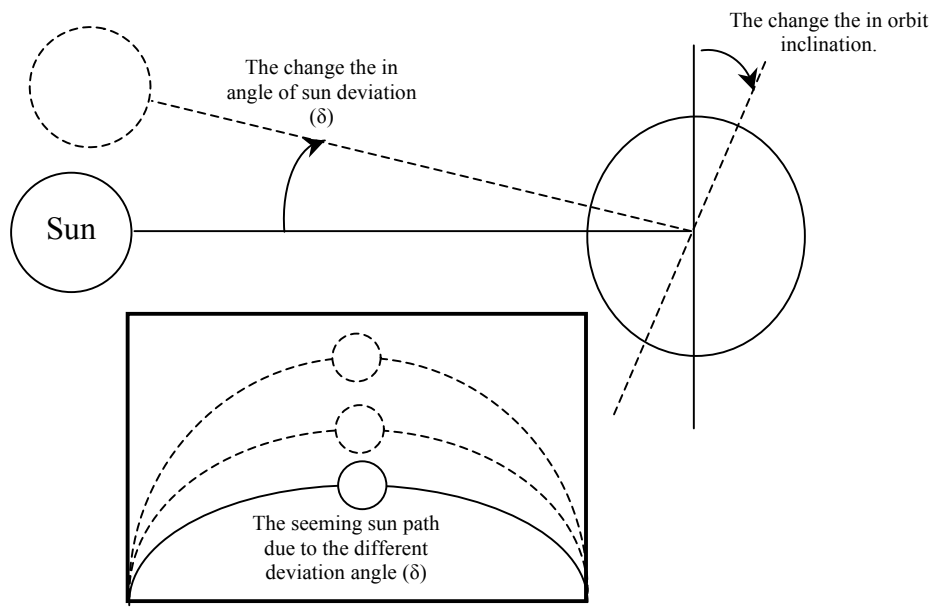

Figure (5) A sketch to the sun seeming deviation angle.

**The geographic latitude angle ($\phi$):** this is the earth point coordinate, it is positive to the north of the Earth, and zero at the equator line.

The shadow of all objects is lengthened words toward the poles of the planet, even if they are in the same latitude.

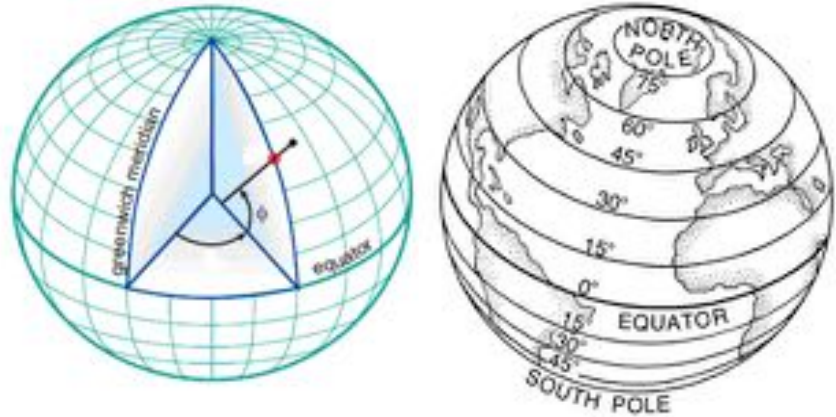

Figure (6) The geographic latitude angle (ϕ).

**The hour angle (ω):** the hour angle measured from the zenith line to the horizon. It is zero at noon, and with positive value along the morning. At the sine shine, it equals $90^o$, and it is distributed equally with the hours of the day half. The same graduating after noon but in negative values.

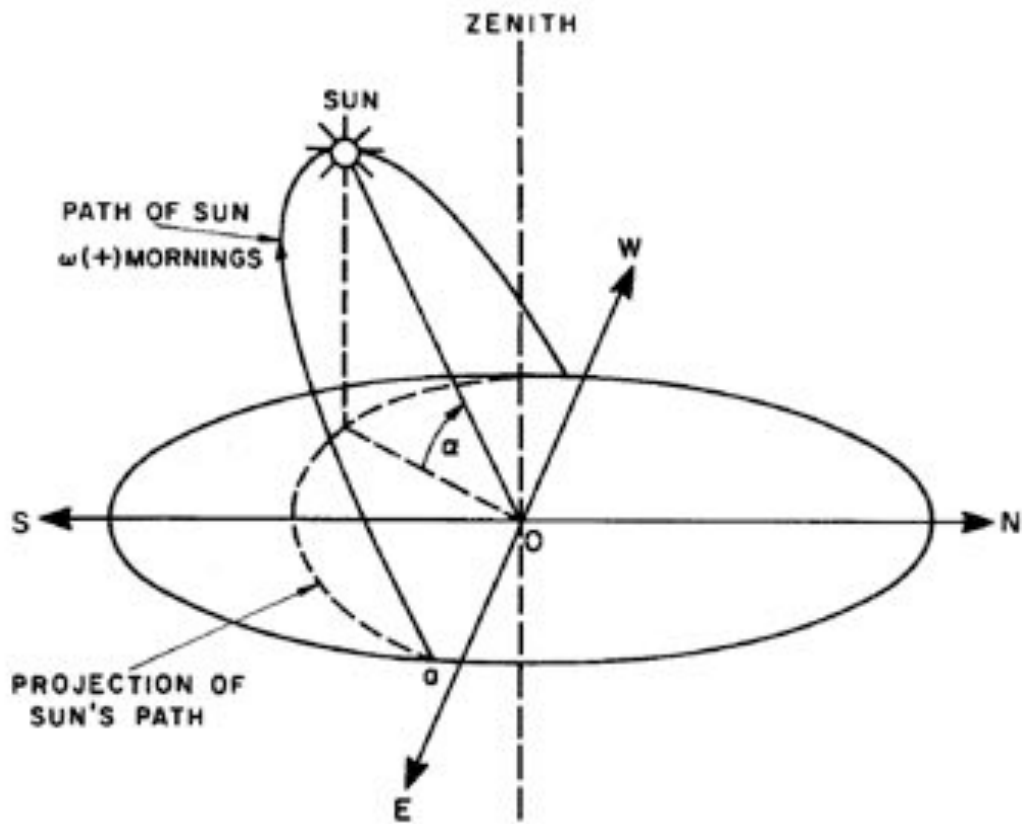

Figure (7) The hour angle (ω).

**The angle of the solar rise (α):** is the angular height of the sun in the sky measured from the horizon (the horizon of the observation with the Earth is the circle obtained from the intersection of the sky of the observer with the Earth). This angle is equal to zero at sunrise and sunset, and 90 when the sun is directly above the observatory. It also may be named solar elevation angle.

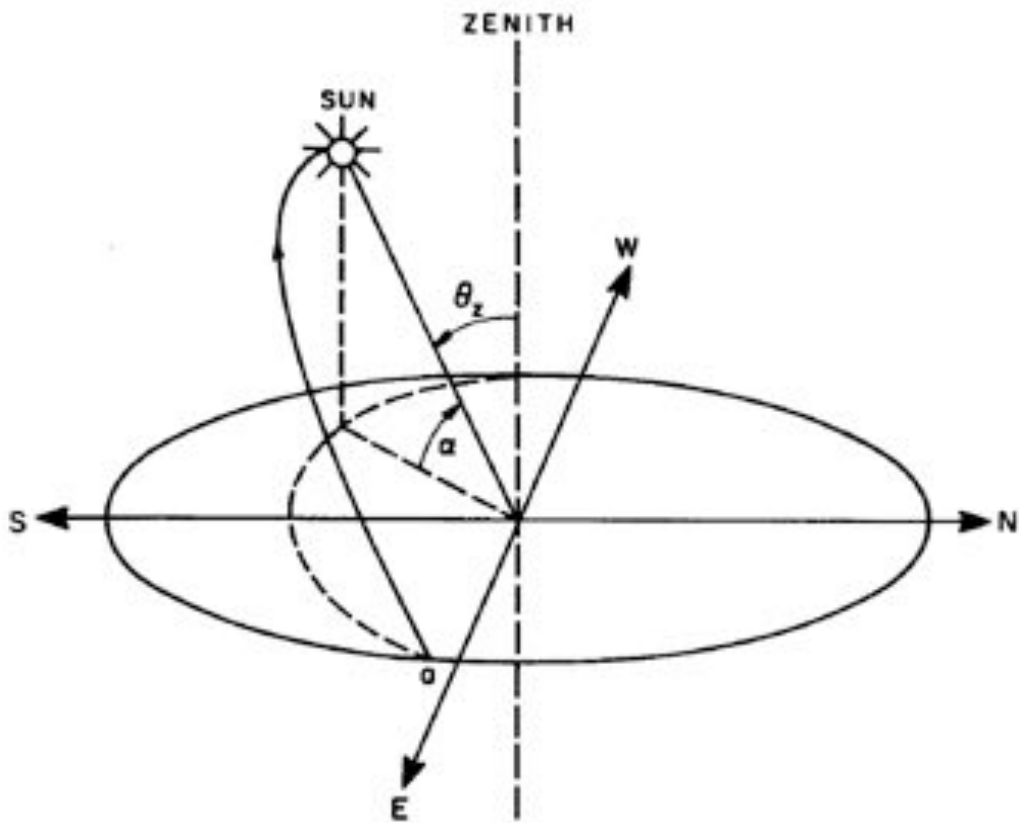

Figure (8) The angle of the solar rise (α).

**The azimuth angle ($\theta_z$):** is the complementary angle of the angle of elevation (α), and measured from the roots, which is after the center of the sun degrees from the point of the head (Zenith), and roughly the angle of the head's head at sunrise or sunset is 90 degrees. It is also named (zenith angle), and can be calculated from the equation below [R. 11, p. 15]:

α= 90- θ$_z$ …………………………………………………(5)

The elevation angle of the relationship is calculated by [R. 11, p.15]:

Cos(θz)=Sin(α)= Cos(φ) × Cos(δ) × Cos(ω) + Sin(φ) × Sin(δ)

…………………………(6)

Now, we can calculate every sun coordinate, or angles depending, and shadow on the number of days for everybody when its orientation and dimensions are known.

## 4. The selected building

We selected a building with known height and orientation to test the accuracy of AutoCAD in addition to the test according to the astronomy equations.

The selected building is a (Qasr Al-Safer hotel and restaurant) in Najaf Government, City, Hai Al- Zehraa. The orientation of it is (32$^o$ 00' 02.49" N) and (44$^o$ 21' 32.88" E).  or (32.000691 N 44.359133 E).

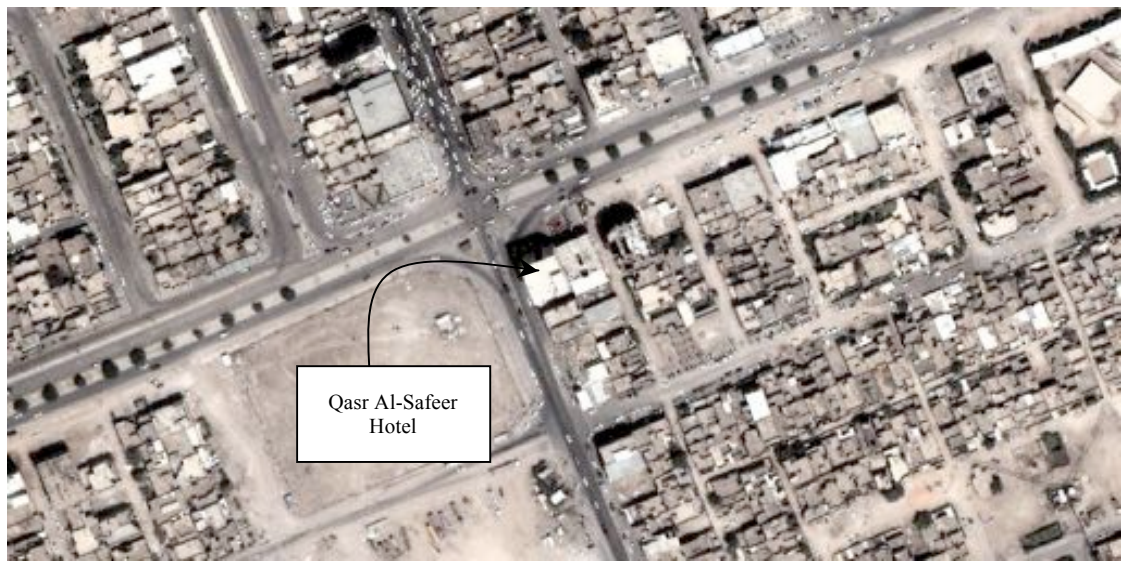

Figure (9) The orientation of Qasr Al-Safeer hotel and restaurant.

The image above is a QuickBird imagery by WGS 84 map projections [R. 14]. And the seen above is imaged at 09:43 a.m. according to the key information file which is attached with the imagery CD.

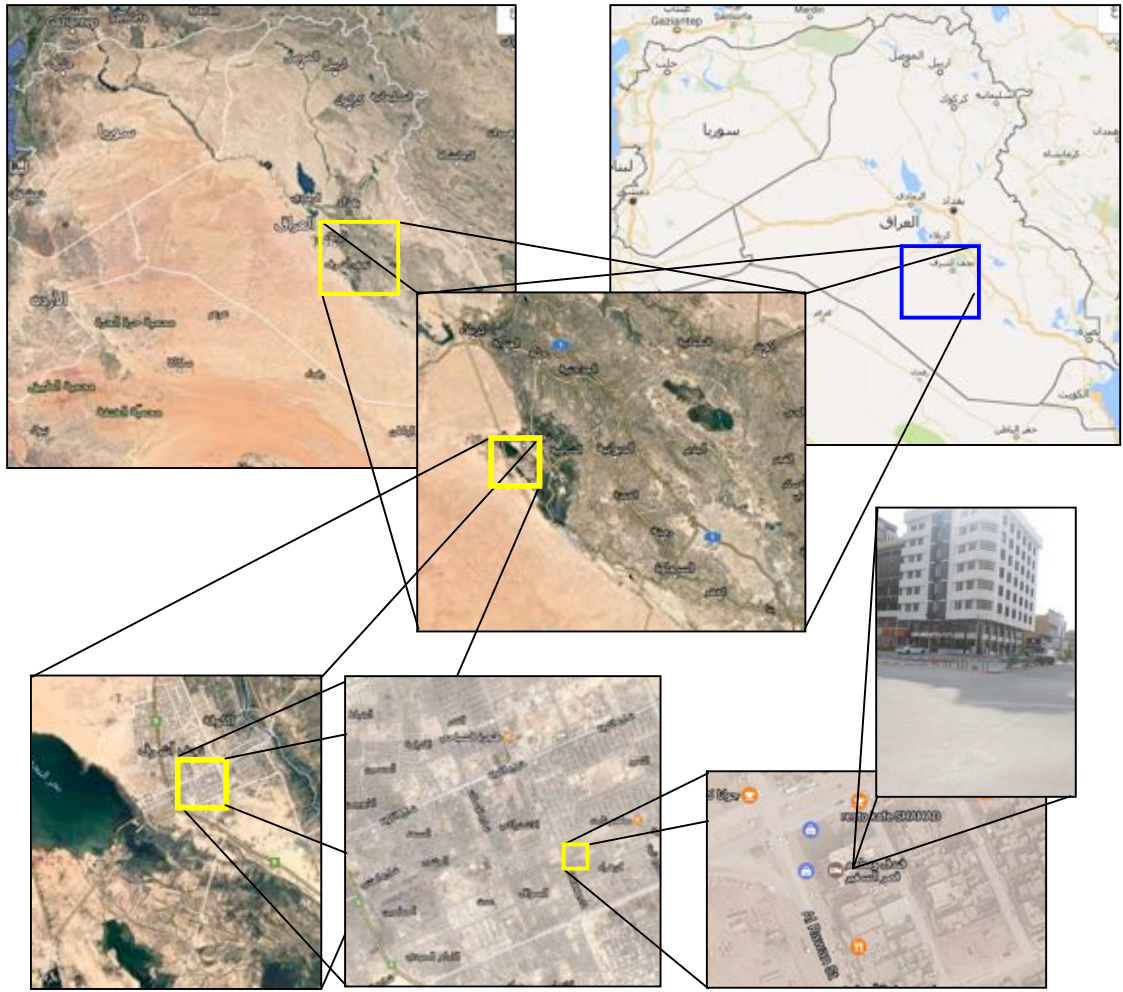

Figure (10) The location of the selected building.

The building is 25.30 m height over the adjacent street level, on the crossing of two streets (Al-Rawan and Al- Ameer), and there are no high buildings around it except at one side. These properties make easy to measure the length of shadow from two sides. By the electronic and magnetic composes, we found the direction of building walls to measure the direction of shadow according to it, by this we need no repeat the field measurements of shadow according to the eastern direction.

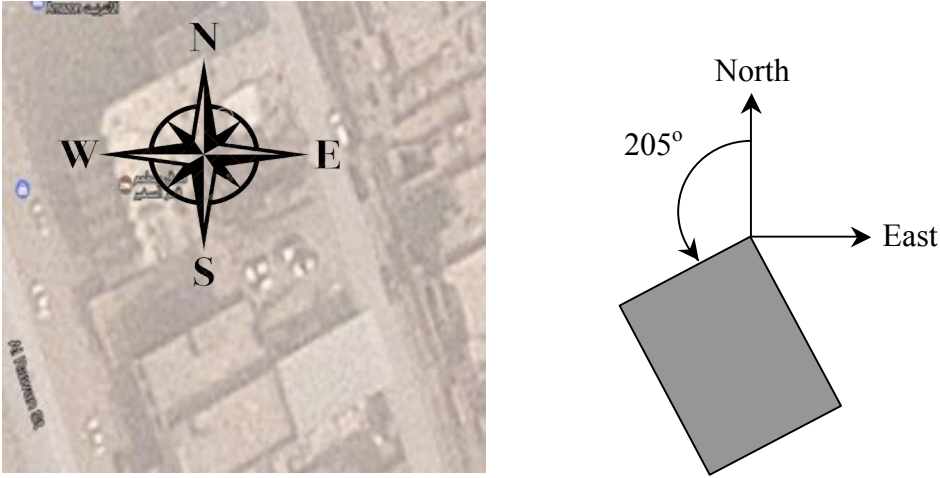

Figure (11) The Direction of the walls of the selected building.

## 5. Calculate the sun orientations during the year

The sun orientation in the sky is calculated from astronomy equations during the years as presented.

For the QuickBird imaging time at 29/8/2006:

The day number ($d_n$) is 241.

The deviation angle ($\delta$) from equation (2)

$\Gamma = 2\pi(d_n-1)/365 = 4.131409517$ rad

$\delta = (0.006918-0.399912\times\cos(4.13141)+0.070257\times\sin(4.13141)-0.006758\times\cos(2\times4.13141) +0.000907\times\sin(2\times4.13141)-0.002697 \times \cos(3\times4.13141) +0.00148\times\sin(3\times4.13141)\times(180/\pi)$

$\delta = 0.168285459$ rad

$= 9.642046551$ degrees

The latitude angle ($\phi$) is + 32.000691 degrees. From the latitude circle.

The hour (solar) angle ($\omega$)= 12:00- 09:43= 02:17 hours = 34.25 degrees

$= 0.711$ rad

The zenith angle ($\theta_z$) from eq. (6)

$\cos(\theta_z)=\sin(\alpha)= \cos(\phi) \times \cos(\delta) \times \cos(\omega) + \sin(\phi) \times \sin(\delta)$

$\qquad = \cos(32) \times \cos(9.642) \times \cos(34.25) + \sin(32) \times \sin(9.642)$

$\qquad = 0.848 \times 0.98587 \times 0.82658975 + 0.52992 \times 0.16749$

$\qquad = 0.7798$

$\theta_z = 38.7577$ degrees

The angle of solar rise ($\alpha$)= 51.24 degrees from eq. (5)

From this angle we can calculate the shadow length:

$\tan(\alpha)$= building height /shadow width (measured from field or image at that date: 29/August)

$\qquad$ = H/ 19.95 m

H= $19.95 \times \tan(51.24227)$= 24.85 m

But from the field the height of building is 25.30 m

So, if we calculate the shadow geometrically from the astronomic equation with known height.

$\tan(\alpha)$= 25.3/shadow length

shadow length= $25.3/\tan(\alpha)$= 20.31 m

It is easy to program these calculations by any programming language. Then we make a table to all shadow length.

Table (1) Shadow length and direction by the astronomy equations (rounded values to two digits).

| date | day number | day angle | delineation angle | latitude angle | hour angle | azimuth angle | solar rise angle | shadow length |
|---|---|---|---|---|---|---|---|---|
| 29-Aug | 241 | 4.13 | 0.17 | 32 | 34.25 | 0.68 | 51.25 | 20.31 |
| 13-Sep | 256 | 4.39 | 0.07 | 32 | 34.25 | 0.74 | 47.49 | 23.19 |
| 28-Sep | 271 | 4.65 | -0.03 | 32 | 34.25 | 0.82 | 43.23 | 26.91 |
| 13-Oct | 286 | 4.91 | -0.13 | 32 | 34.25 | 0.89 | 38.77 | 31.50 |
| 28-Oct | 301 | 5.16 | -0.22 | 32 | 34.25 | 0.97 | 34.43 | 36.90 |

| date | day number | day angle | delineation angle | latitude angle | hour angle | azimuth angle | solar rise angle | shadow length |
|---|---|---|---|---|---|---|---|---|
| 12-Nov | 316 | 5.42 | -0.31 | 32 | 34.25 | 1.04 | 30.61 | 42.76 |
| 27-Nov | 331 | 5.68 | -0.37 | 32 | 34.25 | 1.09 | 27.68 | 48.23 |
| 12-Dec | 346 | 5.94 | -0.40 | 32 | 34.25 | 1.12 | 25.98 | 51.93 |
| 27-Dec | 361 | 6.20 | -0.41 | 32 | 34.25 | 1.12 | 25.70 | 52.57 |
| 11-Jan | 11 | 0.17 | -0.38 | 32 | 34.25 | 1.10 | 26.90 | 49.88 |
| 26-Jan | 26 | 0.43 | -0.33 | 32 | 34.25 | 1.06 | 29.42 | 44.86 |
| 10-Feb | 41 | 0.69 | -0.26 | 32 | 34.25 | 0.99 | 33.00 | 38.96 |
| 25-Feb | 56 | 0.95 | -0.16 | 32 | 34.25 | 0.92 | 37.25 | 33.28 |
| 12-Mar | 71 | 1.20 | -0.06 | 32 | 34.25 | 0.84 | 41.77 | 28.33 |
| 27-Mar | 86 | 1.46 | 0.04 | 32 | 34.25 | 0.76 | 46.19 | 24.27 |
| 11-Apr | 101 | 1.72 | 0.14 | 32 | 34.25 | 0.69 | 50.18 | 21.09 |
| 26-Apr | 116 | 1.98 | 0.23 | 32 | 34.25 | 0.64 | 53.49 | 18.72 |
| 11-May | 131 | 2.24 | 0.31 | 32 | 34.25 | 0.59 | 55.97 | 17.08 |
| 26-May | 146 | 2.50 | 0.37 | 32 | 34.25 | 0.57 | 57.60 | 16.06 |
| 10-Jun | 161 | 2.75 | 0.40 | 32 | 34.25 | 0.55 | 58.45 | 15.54 |
| 25-Jun | 176 | 3.01 | 0.41 | 32 | 34.25 | 0.55 | 58.64 | 15.42 |
| 10-Jul | 191 | 3.27 | 0.39 | 32 | 34.25 | 0.56 | 58.19 | 15.69 |
| 25-Jul | 206 | 3.53 | 0.35 | 32 | 34.25 | 0.57 | 57.06 | 16.39 |
| 9-Aug | 221 | 3.79 | 0.28 | 32 | 34.25 | 0.61 | 55.12 | 17.64 |
| 24-Aug | 236 | 4.05 | 0.20 | 32 | 34.25 | 0.66 | 52.35 | 19.52 |

## 6. The fields measurements

A scale tape and simple instruments achieved the field measurements.

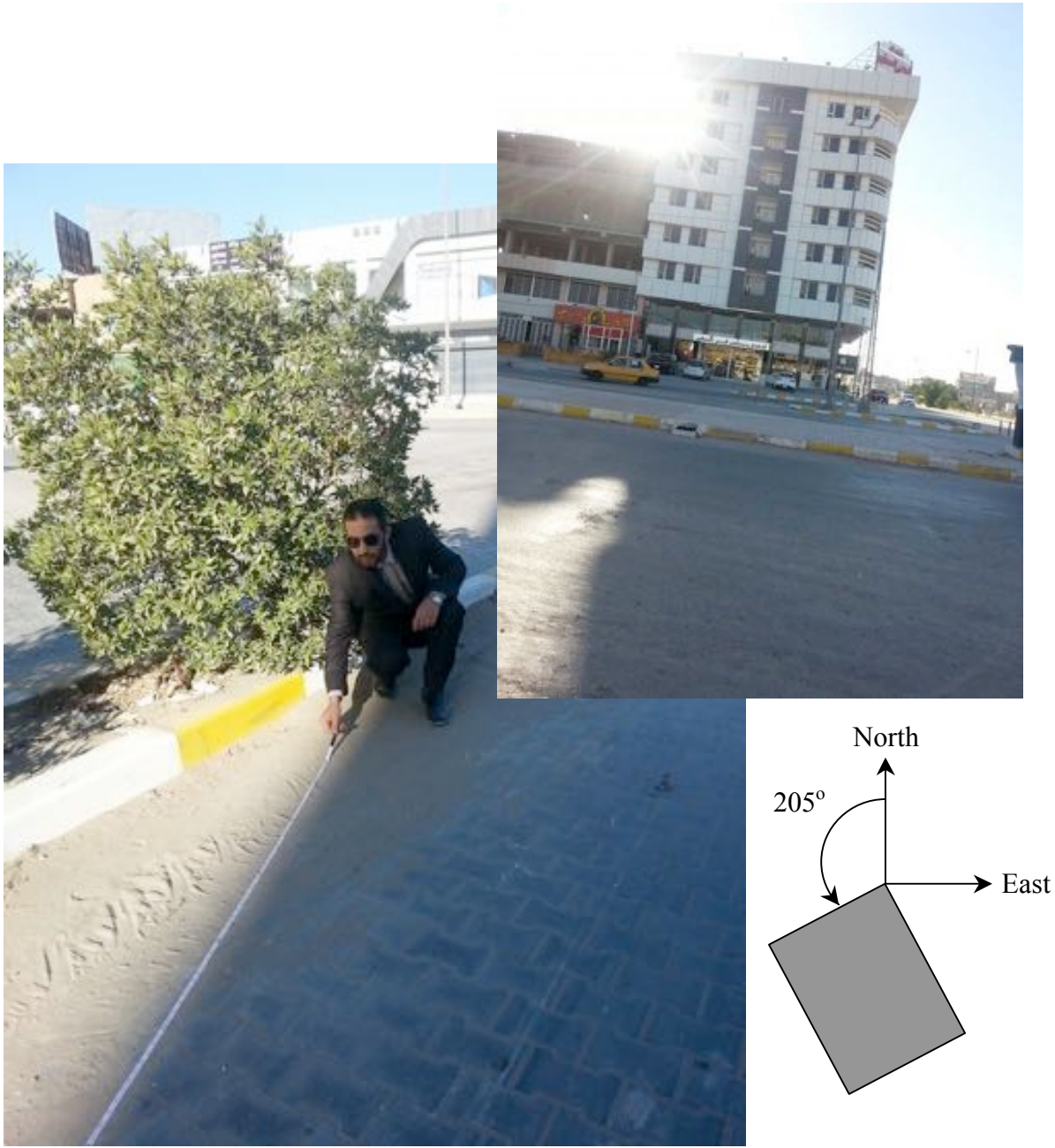

Figure (12) Field measurements of shadow length and direction.

The results are illustrated in the table below:

Table (2) Illustrate the field measurements for the length and direction of building shadow.

| Date | Shadow angle from the North (degrees) | shadow length (m) |
|---|---|---|
| 29-Aug | 51 | 20.20 |
| 13-Sep | 47 | 23.05 |
| 28-Sep | 43 | 26.85 |
| 13-Oct | 38 | 31.45 |
| 28-Oct | 34 | 36.80 |
| 12-Nov | 30 | 42.70 |
| 27-Nov | 27 | 48.15 |
| 12-Dec | 26 | 51.85 |
| 27-Dec | 26 | 52.50 |
| 11-Jan | 27 | 49.80 |
| 26-Jan | 29 | 44.80 |
| 10-Feb | 33 | 38.90 |
| 25-Feb | 37 | 33.20 |
| 12-Mar | 42 | 28.25 |
| 27-Mar | 46 | 24.20 |
| 11-Apr | 50 | 21.00 |
| 26-Apr | 53 | 18.65 |
| 11-May | 55 | 17.00 |
| 26-May | 57 | 16.00 |
| 10-Jun | 58 | 15.50 |
| 25-Jun | 58 | 15.35 |
| 10-Jul | 58 | 15.60 |
| 25-Jul | 57 | 16.30 |
| 9-Aug | 55 | 17.60 |
| 24-Aug | 52 | 19.50 |

Also we measured the shadow length and direction in a spread days during the year to compare with AutoCAD results.

## 7. The AutoCAD package

The AutoCAD program is a commercial computer aided drawing (designing). Designed by Autodesk Company, was released firstly in 1982 under MS-DOS environment and running on microcomputers. Then it developed with the development of hardware and software apparatuses, and became the most used software for engineers and designer. The newer version of AutoCAD which are under Windows Operating Systems have rendering equipment, one of them is the lightening and orientation tools.

## 8. A 3d Model for our site

We depended AutoCAD Version 2015 to design a model similar to our building to check our work and measurements to the shadow of our building that AutoCAD simulating it. The model is a box with 26.3 m height upon a plate of 1 m height to represent the height of the hotel building.

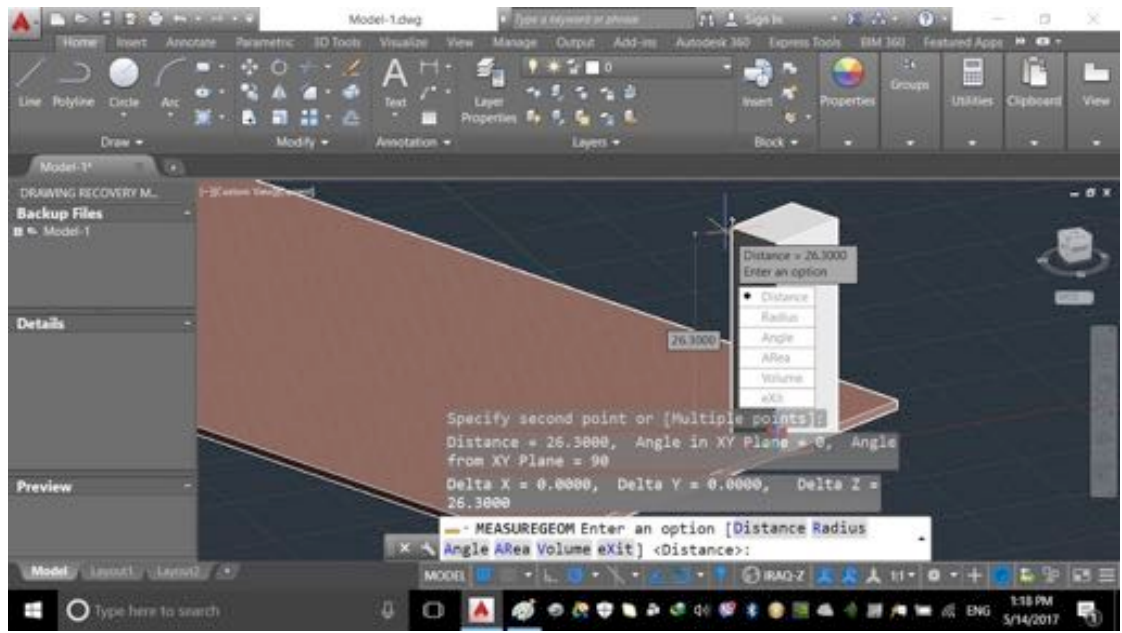

Figure (13) A 3D model by AutoCAD 2015.

Then, we defined the location of the building from Map tool, set the latitude and longitudinal coordinates, the local time, and a direction at 75 from the north as shown below.

ne of the building.

afte          the  default

at  the  maging  date  of  the

Figure (15) Set the sun status and the date at the year.

The render tool is very important tool, the shadow does not appear unless render the model. We were render the scene region to show the shadow extending.

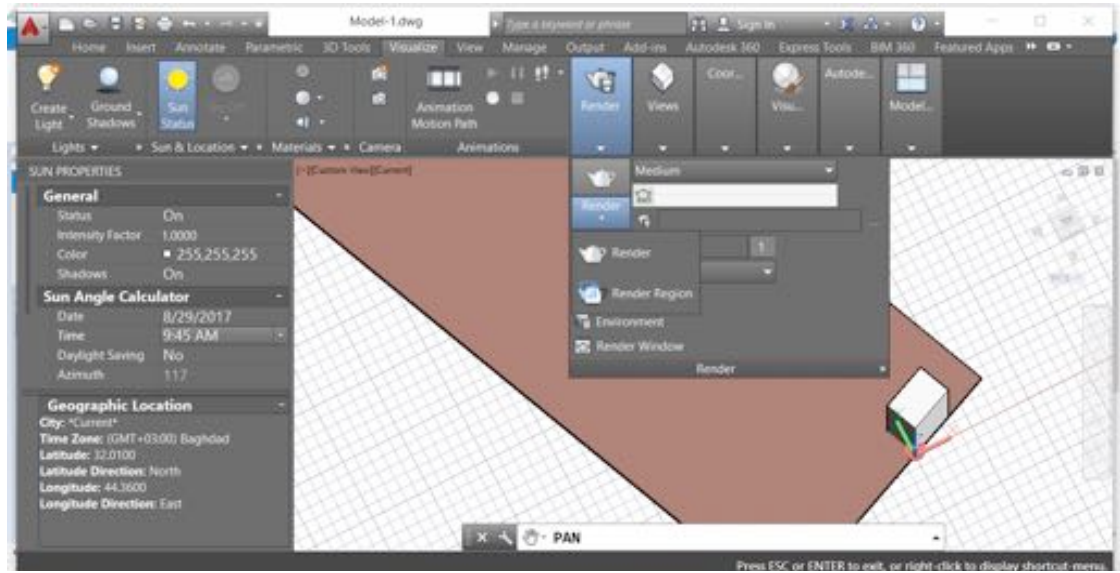

Figure (16) The Rendering tool.

In rendered scene we need to use the measure tool to find the length (Distance) and (Angle) of the shadow in the rendered scene.

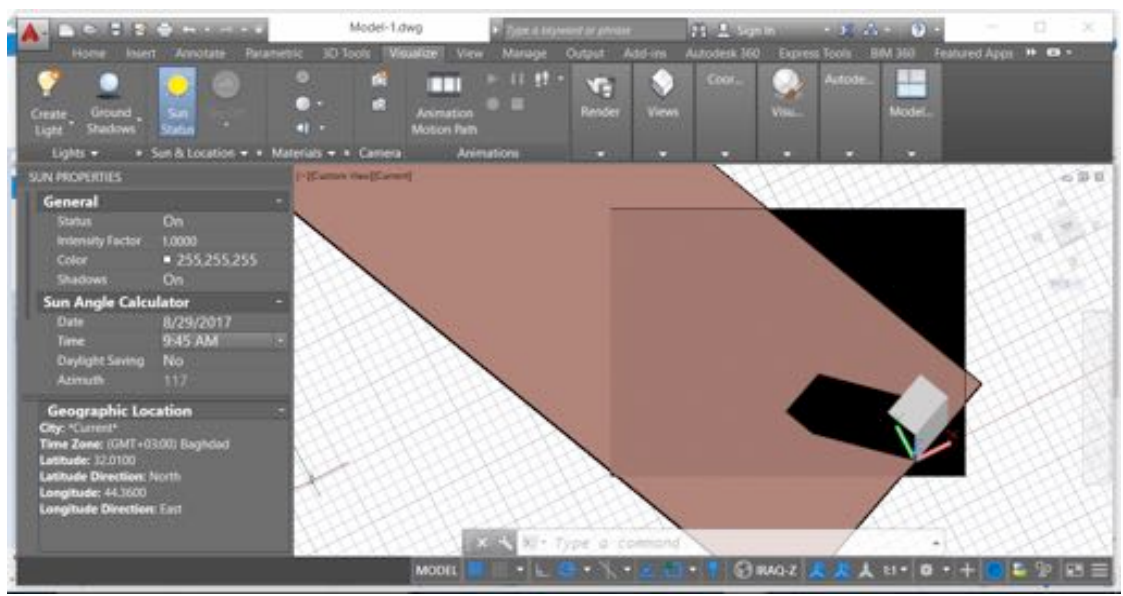

Figure (17) The measure tool in the Utilities.

By the help of snap tool to target the edge of the box, and select the dedicated shadow end by mouse clicking.

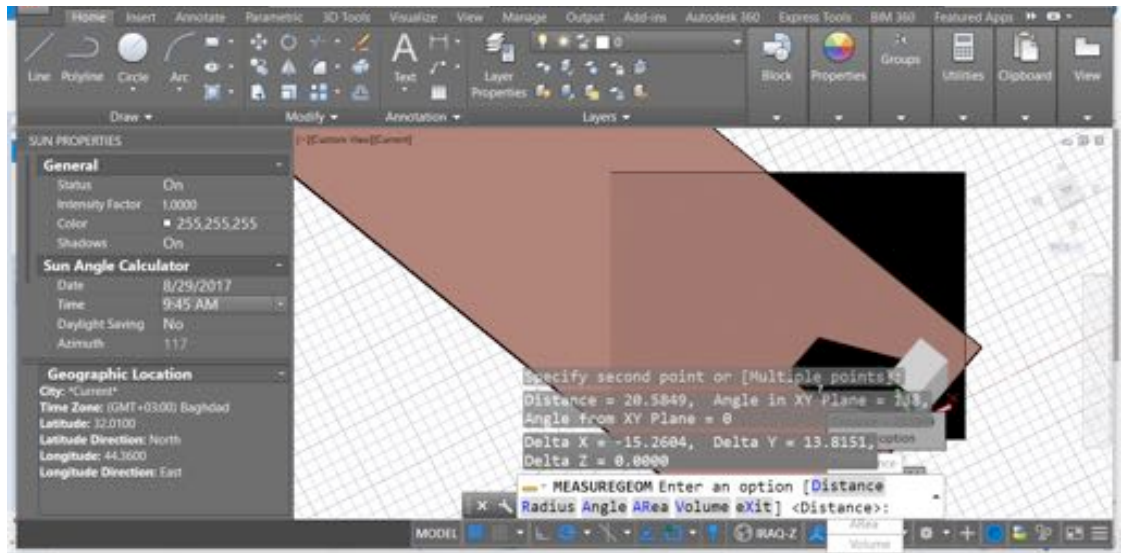

Figure (18) measurements of the distance and angle of the shadow from the east side.

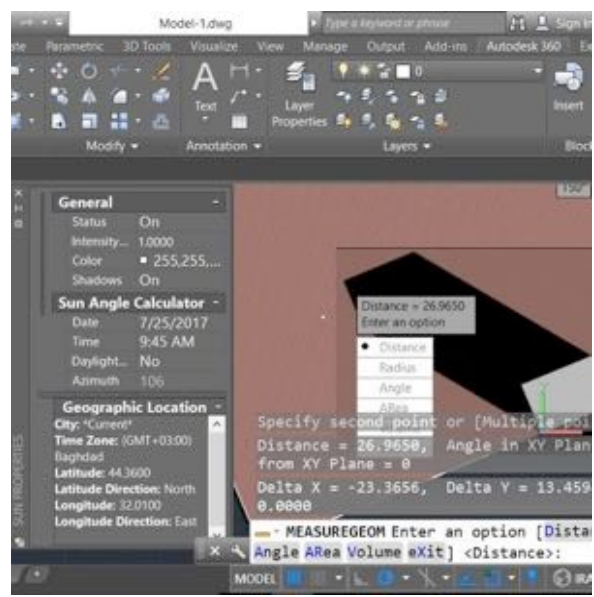

Figure (19) changing the date and measurement the shadow length and angle during the year.

We record the shadow length and direction for selected dates during the year to compare the AutoCAD model with the astronomy calculations and field measurements.

Table (3) Calculations of the AutoCAD model.

| Date | AutoCAD shadow length | AutoCAD shadow direction | Angle from North |
|------|------------------------|---------------------------|-------------------|
| 29-Aug | 34.4394 | 139 | 49 |
| 13-Sep | 39.2743 | 134 | 44 |
| 28-Sep | 47.1740 | 129 | 39 |
| 13-Oct | 54.6891 | 124 | 34 |
| 28-Oct | 66.7774 | 120 | 30 |
| 12-Nov | 85.5406 | 118 | 28 |
| 27-Nov | 105.6931 | 117 | 27 |
| 12-Dec | 131.0835 | 117 | 27 |
| 27-Dec | 145.5041 | 118 | 28 |
| 11-Jan | 137.6249 | 121 | 31 |
| 26-Jan | 115.6797 | 123 | 33 |
| 10-Feb | 89.8401 | 126 | 36 |
| 25-Feb | 68.7173 | 129 | 39 |
| 12-Mar | 53.0068 | 132 | 42 |
| 27-Mar | 42.8531 | 135 | 45 |
| 11-Apr | 35.2756 | 139 | 49 |
| 26-Apr | 30.4343 | 142 | 52 |
| 11-May | 26.946 | 146 | 56 |
| 26-May | 25.2833 | 150 | 60 |
| 10-Jun | 24.0221 | 152 | 62 |
| 25-Jun | 24.2047 | 153 | 63 |
| 10-Jul | 25.3407 | 153 | 63 |
| 25-Jul | 26.965 | 150 | 60 |
| 9-Aug | 29.3303 | 146 | 56 |
| 24-Aug | 32.8399 | 141 | 51 |

## 9. Comparing the Results

It is clear that the shadow in the AutoCAD model seems longer that the shadow length that measured in field or from the astronomy equations.

The figure below is describing the difference between the results, which refer to the high difference, about 45%.

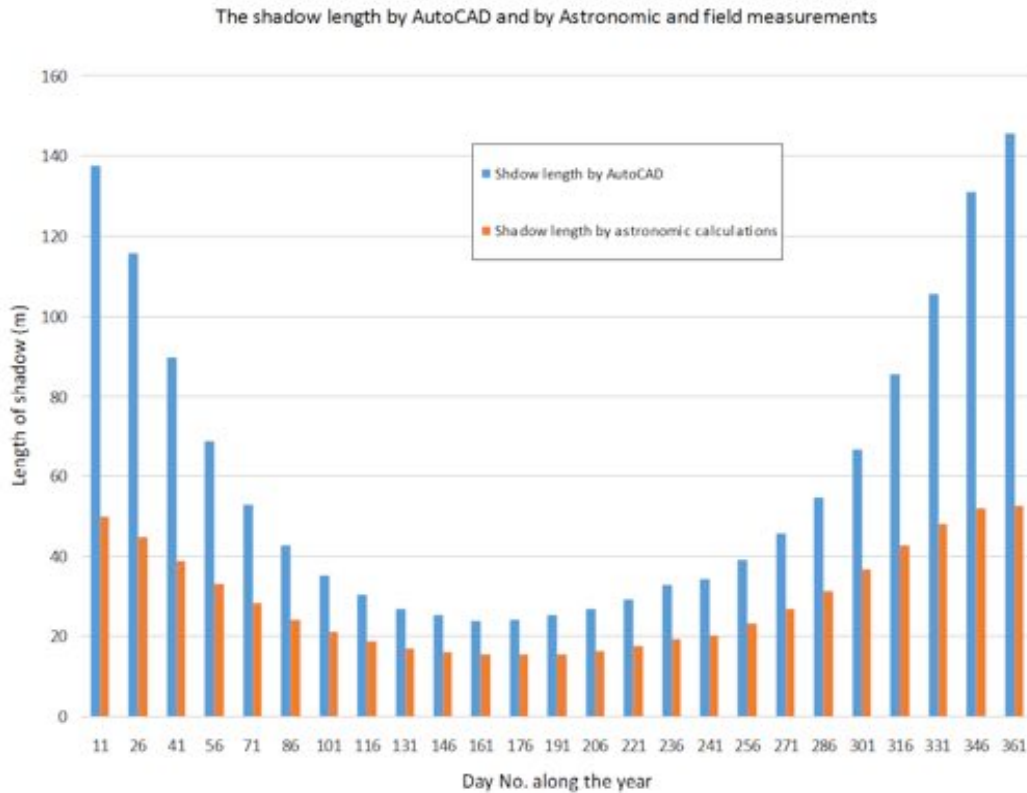

Figure (20) The length of shadow is longer than the real.

The direction of the shadow also differs between the AutoCAD model, the real measurements, and the astronomy calculations.

## 10. The Results and Conclusions

The final decision according to the comparing of the results, that the AutoCAD sun status and sun lightening tools is inaccurate tools, and need to depend the astronomy equations.

We tested these tools in our site and coordinates, but we suggest to test these work for more places to find the weak point in the AutoCAD lightening and rendering tools over the world.

We were used AutoCAD 2015 and 2016 version, but the sun calculation lightening tool remain same until 2018 version.

Both AutoCAD and 3D Studio Max used the dynamic linking library of the graphic accelerator Direct3D of DirectX, which has the programming and equations those represent the bass of Autodesk products design, so we expect the same errors in the other product, 3D Studio Max and other Autodesk products.